x
\documentclass[twocolumn,usenatbib]{aastex62}
\usepackage{lipsum}
\usepackage{totcount}
\usepackage{comment}
\usepackage{floatrow}
\usepackage[caption=false]{subfig}
\usepackage{amsmath}

\graphicspath{{./}{}}

\submitjournal{ApJL}

\shorttitle{Detection of ionized calcium in the atmosphere of KELT-9b}
\shortauthors{Turner J.D, de Mooij E.J.W., Jayawardhana R., et al.}

\begin{document}

\title{Detection of ionized calcium in the atmosphere of the ultra-hot Jupiter KELT-9b}

\correspondingauthor{Jake D. Turner}
\email{astrojaketurner@gmail.com, jaketurner@cornell.edu}

\author[0000-0001-7836-1787]{Jake D. Turner}
\affil{Department of Astronomy and Carl Sagan Institute, Cornell University, Ithaca, New York 14853, USA}

\author[0000-0001-6391-9266]{Ernst J. W. de Mooij}
\affil{Astrophysics Research Centre, Queen's University Belfast, Belfast BT7 1NN, UK}
\affil{School of Physical Sciences and Centre for Astrophysics $\&$ Relativity, Dublin City University, Glasnevin, Dublin, Ireland}

\author{Ray Jayawardhana}
\affil{Department of Astronomy, Cornell University, Ithaca, New York 14853, USA}

\author{Mitchell E. Young}
\affil{Space Research Institute, Austrian Academy of Sciences, Schmiedlstrasse 6, A-8042 Graz, Austria}

\author[0000-0003-4426-9530]{Luca Fossati}
\affil{Space Research Institute, Austrian Academy of Sciences, Schmiedlstrasse 6, A-8042 Graz, Austria}

\author{Tommi Koskinen}
\affil{Lunar and Planetary Laboratory, University of Arizona, 1629 East University Boulevard, Tucson, AZ 85721-0092, USA}

\author{Joshua D. Lothringer}
\affil{Lunar and Planetary Laboratory, University of Arizona, 1629 East University Boulevard, Tucson, AZ 85721-0092, USA}
\affil{Department of Physics $\&$ Astronomy, Johns Hopkins University, Baltimore, MD 21218, USA}

\author{Raine Karjalainen}
\affil{Isaac Newton Group of Telescopes 321, Apartado de Correos, Santa Cruz de La Palma, E-38700, Spain}
\affil{Instituto de Astrof\'{i}sica de Canarias, c/ V\'{i}a L$\acute{a}$ctea s/n E-38205 La Laguna, Tenerife, Spain} 

\author{Marie Karjalainen}
\affil{Instituto de Astrof\'{i}sica de Canarias, c/ V\'{i}a L$\acute{a}$ctea s/n E-38205 La Laguna, Tenerife, Spain} 



\begin{abstract}

\noindent With a day-side temperature in excess of 4500K, comparable to a mid-K-type star, KELT-9b is the hottest planet known. Its extreme temperature makes KELT-9b a particularly interesting test bed for investigating the nature and diversity of gas giant planets. We observed the transit of KELT-9b at high spectral resolution (R$\sim$94,600) with the CARMENES instrument on the Calar Alto 3.5-m telescope. Using these data, we detect for the first time ionized calcium (Ca{\sc ii} triplet) absorption in the atmosphere of KELT-9b; this is the second time that Ca{\sc ii} has been observed in a hot Jupiter. Our observations also reveal prominent H$\alpha$ absorption, confirming the presence of an extended hydrogen envelope around KELT-9b. We compare our detections with an atmospheric model and find that all four lines form between atmospheric temperatures of 6100\,K and 8000\,K and that the Ca{\sc ii} lines form at pressures between 10 and 50\,nbar while the H$\alpha$ line forms at a lower pressure ($\sim$6\,nbar), higher up in the atmosphere. The altitude that the core of H$\alpha$ line forms is found to be $\sim$1.4 R$_{p}$, well within the planetary Roche lobe ($\sim$1.9\,R$_{p}$). Therefore, rather than probing the escaping upper atmosphere directly, the H$\alpha$ line and the other observed Balmer and metal lines serve as atmospheric thermometers enabling us to probe the planet's temperature profile, thus energy budget.

\end{abstract}



\section{Introduction} \label{sec:intro}
Characterizing exoplanetary atmospheres provides valuable insights into their composition, climate, evolution, and habitability. Transiting exoplanets are of particular interest because they offer the opportunity to obtain atmospheric transmission spectra \citep{Charbonneau2000}. Transit observations have been conducted in both broad-band spectral windows (e.g. \citealt{Sing2016}) and at high spectral resolution (R$\sim$50,000--100,000; e.g. \citealt{Snellen2010}). Observing such a spectrum makes it possible to detect individual atomic lines (e.g. \citealt{Charbonneau2002}; \citealt{VidalMadjar2003}; \citealt{Redfield2008}; \citealt{Jensen2012}; \citealt{Deibert2019}; \citealt{Casasayas2019}; \citealt{Nortmann2018}; \citealt{Cauley2019}; \citealt{Hoeijmakers2018}) and molecular species (e.g \citealt{Snellen2010}; \citealt{Brogi2012}; \citealt{Kreidberg2015}; \citealt{Sing2016}; \citealt{Birkby2017}; \citealt{Esteves2017}; \citealt{Turner2016b, Turner2017}; \citealt{Evans2018}) in the planetary atmosphere. Additionally, extended hydrogen envelopes have been observed around several hot Jupiters (e.g. \citealt{VidalMadjar2003}; \citealt{Lecavelier2012}). Measuring the slope of the transmission spectra can be used to infer the presence of atmospheric clouds or haze (e.g. \citealt{Pont2008}; \citealt{Bean2010}; \citealt{Knutson2014}; \citealt{Kreidberg2014}; \citealt{Turner2013a,Turner2016b}; \citealt{Sing2016}) and scattering particles (e.g. \citealt{Pont2008}; \citealt{Sing2011}; \citealt{Turner2016b}; \citealt{Mallonn2017}). 


KELT-9b orbits a A0V/B9V star at a distance of 0.034 AU (\citealt{Gaudi2017}) and is the hottest known gas giant with a T$_{\text{eff}}$ $\approx$ 4500 K (\citealt{Hooton2018,Wong2019,Mansfield2019}). Due to the high temperature of the planet \cite{Kitzmann2018} predicted that iron absorption lines and several near-ultraviolet resonance lines should be observable in KELT-9b's transmission spectra. Neutral and ionized iron in its atmosphere have subsequently been detected in multiple HARPS-N observations (\citealt{Hoeijmakers2018,Hoeijmakers2019}). Also, \citet{Yan2018} reported an extended hydrogen atmosphere ($\sim$\,1.64\,$R_{p}$) using H$\alpha$ observations with the CARMENES spectrograph. The H$\alpha$ detetion was confirmed by \citet{Cauley2019}, who observed variability in the in-transit absorption light curve suggesting that the transiting gas is not uniform. Using high-resolution optical spectra from HARPS-N during four separate KELT-9b transits \citet{Borsa2019} observed the atmospheric Rossiter-McLaughlin effect. In total, 14 atomic species have been reported previously in KELT-9b's atmosphere (\citealt{Yan2018}; \citealt{Cauley2019};  \citealt{Hoeijmakers2018,Hoeijmakers2019}).  

Here, we present observations of KELT-9b using the CARMENES spectrograph on the 3.5-m Calar Alto telescope. With these data, we detect ionized calcium (Ca{\sc ii}) for the first time in its atmosphere. Ca{\sc ii} is predicted to be in hot Jupiter atmospheres (e.g. \citealt{Turner2016a}) and has been observed in the atmospheres of the super-Earth 55 Cnc e (\citealt{RiddenHarper2016}) and the hot Jupiter KELT-20b (\citealt{Casasayas2019}). Our data also reveal prominent H$\alpha$ absorption, confirming the presence of an extended hydrogen envelope  (\citealt{Yan2018} and \citealt{Cauley2019}).

\section{Observations and Data Reduction}
We observed the transit of KELT-9b on June 6-7, 2018 from 21:34--3:53 UT with the CARMENES instrument (\citealt{Quirrenbach2016}) on the 3.5-m Calar Alto telescope. For the main analysis in the paper, we use data from the visible spectrograph which covers the wavelength range 5200--9700 \AA, over 61 spectral orders with a resolution of $\sim$94,600. To search for helium we also use the near-infrared (near-IR) spectrograph which has a wavelength range of 9600--17100 \AA, over 28 spectral orders at a resolution of $\sim$80,400. Exposure times of 113 and 121 seconds were used for the optical and near-IR spectrographs, respectively. Over the entire observation 140 exposures were taken. The airmass varied from 2.10--1.07 during the night. 

\begin{figure*}[]
    \centering
\begin{tabular}{c}
   \includegraphics[width=0.90\textwidth,page=1]{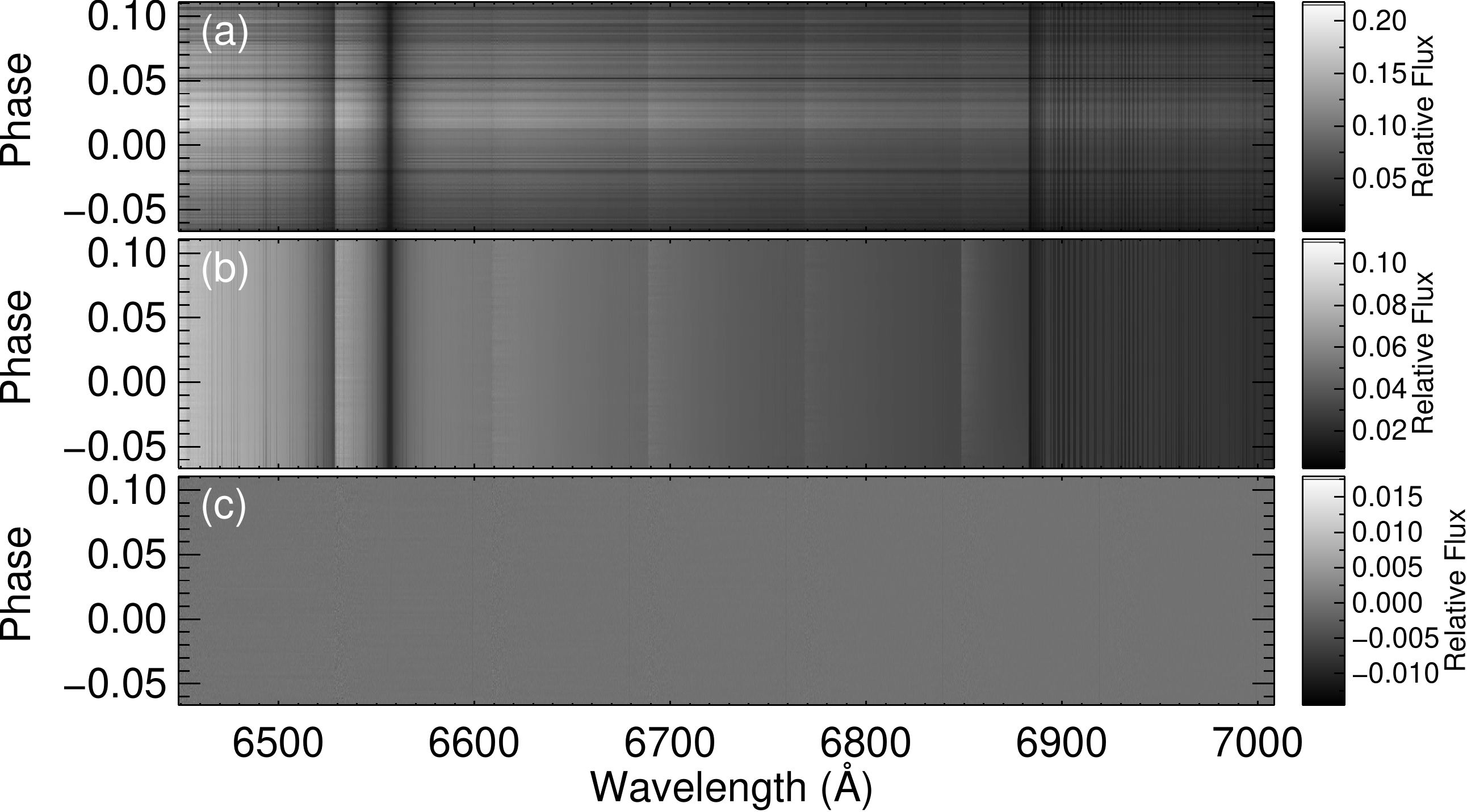} \\
\end{tabular}   
    \caption{Example of the analysis steps done on the CARMENES data. panel (a): Reduced data from the \texttt{CARACAL v2.01} pipeline. panel (b): Data after blaze correction. panel (c) Data after blaze correction and the telluric and stellar lines and other systematics have been removed using \texttt{SYSREM}.  }
    \label{fig:Steps}
\end{figure*}

Standard data reduction was performed with the CARMENES Reduction and Calibration (\texttt{CARACAL v2.01}) pipeline (\citealt{Caballero2016}). This includes bias correction, flat-fielding, cosmic ray removal, and wavelength calibration. An example of the result of this standard reduction can be found Figure in \ref{fig:Steps}a. An additional step was taken to ensure more accurate wavelength calibration. We used a telluric transmission spectrum taken from the ESO Cerro Paranal Advanced Sky Model (\citealt{Moehler2014}) and cross-correlated it with each order of our data. The wavelengths of each order were then linearly shifted by the value that optimized the cross correlation. We found offsets between 0.13--0.2 \AA\ depending on the order. Next, we performed a blaze correction (Section \ref{sec:blaze}) on the data and then removed the telluric and stellar lines and other systematics from the data using \texttt{SYSREM} (Section \ref{sec:sysrem}). 

\subsection{Blaze Normalization} \label{sec:blaze}
As a first step, we correct each order individually by its blaze function. To start, we divide the first science spectrum by the rest of the frames in order to divide out time-invariant changes. Next, each order in this normalized data was binned where outliers in each bin were masked out using the \texttt{PATROL} sigma thresholding technique (\citealt{Turner2019}) and this binned data was fit to a polynomial. Finally, we divided each order in the data from the \texttt{CARACAL v2.01} pipeline by this polynomial fit to remove the blaze function. An example of the blaze corrected data can be found in Figure \ref{fig:Steps}b. 

\subsection{Telluric and stellar subtraction using \texttt{SYSREM}} \label{sec:sysrem}
Next, the telluric and stellar lines and other systematics were removed using \texttt{SYSREM} (\citealt{Tamuz2005}) from the blaze-corrected data. This algorithm has been used extensively to study exoplanet atmospheres (e.g. \citealt{Birkby2017}; \citealt{Esteves2017}; \citealt{Deibert2019}). \texttt{SYSREM} interactively identifies and subtracts systematics (e.g. telluric lines) that are time-invariant while preserving the Doppler-shifted exoplanet signal. Five iterations of \texttt{SYSREM} were ran where each order was treated as a separate ``light curve". If we run between one and five iterations with SYSREM, we obtain the same results within 1$\sigma$ for the absorption profiles and subsequent physical information (see Figure \ref{fig:sysrem}a in Appendix \ref{app:SYSREM}). \citet{Birkby2017} showed that a small part of the planet signal is removed after a few iterations of \texttt{SYSREM}. Therefore, to minimize this effect we use the number of iterations that maximized the S/N in our detection. An example of the data after \texttt{SYSREM} was performed can be found in Figure \ref{fig:Steps}c.

To check the reliability of the data reduction with \texttt{SYSREM}, we also corrected the data using the telluric reduction method described in \citet{Brogi2012} and in Appendix \ref{app:SYSREM}. Using this method, we find consistent results within 1$\sigma$ with \texttt{SYSREM} for the absorption line profiles and widths. A comparison of the H$\alpha$ and Ca-II absorption profiles produced using both methods can be found in panels b-c in Figure \ref{fig:sysrem} in Appendix \ref{app:SYSREM}.

\subsection{Correcting for the Rossiter-McLaughlin effect}
Finally, we removed the  stellar Rossiter–McLaughlin (RM) effect. We modeled these effects using the technique outlined in \citet{deMooij2017} and detailed below. This model was subtracted from the data to get the final spectrum for each order. 

\begin{table*}[!htb]
    \centering
        \caption{Summary of atmosphere line absorption parameters. Column 1: species detected. Column 2: vacuum wavelength of the species (air wavelength). Column 3: transit depth of the absorption profile. Column 4: center velocity of the absorption profile in the rest frame of the planet. Column 5: standard deviation derived from the the absorption profile assuming a Gaussian profile. Column 6: full width at half maximum of the absorption profile derived using $\sigma$ assuming a Gaussian profile. }
    \begin{tabular}{cccccc}
    \hline
    Species &  Wavelength  &  Depth ($\%$)  & $\nu_{cen}$  & $\sigma$ &  FWHM     \\
            & (\AA)       & ($\%$)         & km $s^{-1}$  & km $s^{-1}$ &km $s^{-1}$  \\
    \hline
    H$\alpha$ & 6564.59 (6562.79)       & 1.02$\pm$0.09              & -2.10$^{+1.79}_{-1.82}$ &16.3$^{+1.7}_{-1.5}$        &38.3$^{+4.0}_{-3.5}$  \\
    Ca{\sc ii}   & 8500.35 (8498.02)    & 0.40$\pm$0.05              & -1.76$^{+1.47}_{-1.43}$ &9.8$^{+1.3}_{-1.1}$         &23.1$^{+3.0}_{-2.7}$  \\ 
    Ca{\sc ii}   & 8544.44 (8542.09)    & 0.51$\pm$0.05              & 1.89$^{+1.24}_{-1.38}$  &10.0$^{+1.1}_{-1.3}$        &23.6$^{+2.7}_{-3.1}$  \\
    Ca{\sc ii}   & 8664.52 (8662.14)    & 0.48$\pm$0.06              & 1.99$^{+1.34}_{-1.44}$  & 8.3$^{+1.1}_{-1.0}$        &19.6$^{+2.6}_{-2.4}$ \\
    \hline
    \end{tabular}
    \label{tb:summary}
\end{table*}

The stellar disk was modeled using a regular grid and has a radius of 510 pixels.  The stellar spectrum used for each pixel was obtained from the \texttt{Spectroscopy Made Easy (SME)} software package (\citealt{Piskunov2017}). The stellar parameters used for KELT-9 in the model are $T_{\text{eff}}$ = 10,170 K, $\log(g)$~=~4.093, [Fe/H] = -0.03, and VsinI = 111.4 km $s^{-1}$ \citep{Gaudi2017}. The spectra were generated at 25 different limb angles ($\mu = \cos\theta$) and linearly interpolated in $\mu$ for all other limb angles, to ensure that our model takes the (differential) limb-darkening effects into account. We assume that the star rotates as a solid body, and ignore the effects of gravity darkening. The planet was modeled as an opaque disk in a circular orbit with the planetary parameters (radius, semi-major axis, orbital period, impact parameter and projected spin-orbit misalignment angle) taken from \citet{Gaudi2017}. We note that when modeling the RM effect, we use the R$_p$/R$_*$ from \cite{Gaudi2017}, and do not consider the impact of the excess absorption in the planet's atmosphere at different wavelengths. This will lead to a slight underestimate of the RM amplitude in lines with strong planetary absorption, and thereby a small underestimate of the planetary absorption. After generating the models for each of the frames, we normalized them, and subsequently removed the out-of-transit model in order to obtain the RM effect. We note that it is crucial to normalize the models, as the data are effectively continuum normalized by the blaze-correction. This normalisation will lead to a slight increase in the depth of the line for the parts not occulted by the planet.

\section{Results and Discussion}


We searched for absorption due to individual spectral lines using the final dynamic spectrum after blaze normalization, removing the telluric and stellar lines, and correcting for the RM effect. The composite absorption spectra were generated by correcting for the systemic velocity of the system and then shifting each individual observation into the rest frame of the planet. We used the planetary parameters from \citet{Gaudi2017}, setting the radial velocity semi-amplitude K$_{p}$ = 276 m s$^{-1}$ and the systemic velocity V$_{r}$ = -19.819. We note that the K$_{p}$ values in the literature differ by as much as 5$\sigma$, so care must be taken when looking for offsets in line profiles due to atmospheric winds. To obtain the final composite absorption profiles, we averaged all the data between the second and third transit contact points excluding the region affected by the Doppler shadow.

\begin{figure*}[!htb]
    \centering
    \begin{tabular}{cc}
    \includegraphics[width=0.5\textwidth,page=2]{20180616_Model_H-Alpha.pdf} &
     \includegraphics[width=0.45\textwidth,page=1]{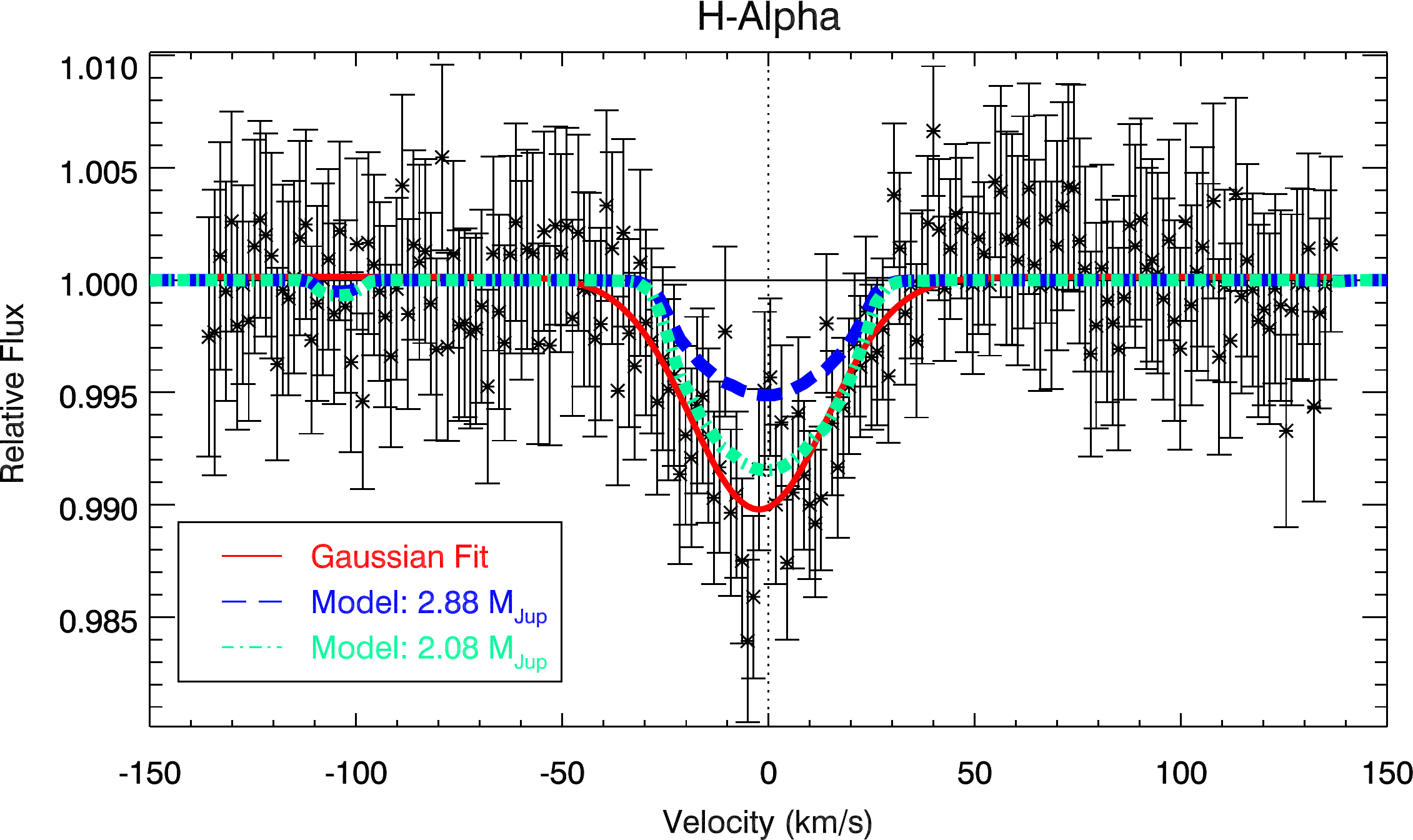}\\
     \includegraphics[width=0.5\textwidth,page=2]{20180616_Model_CaII-8500.pdf} &
     \includegraphics[width=0.45\textwidth,page=1]{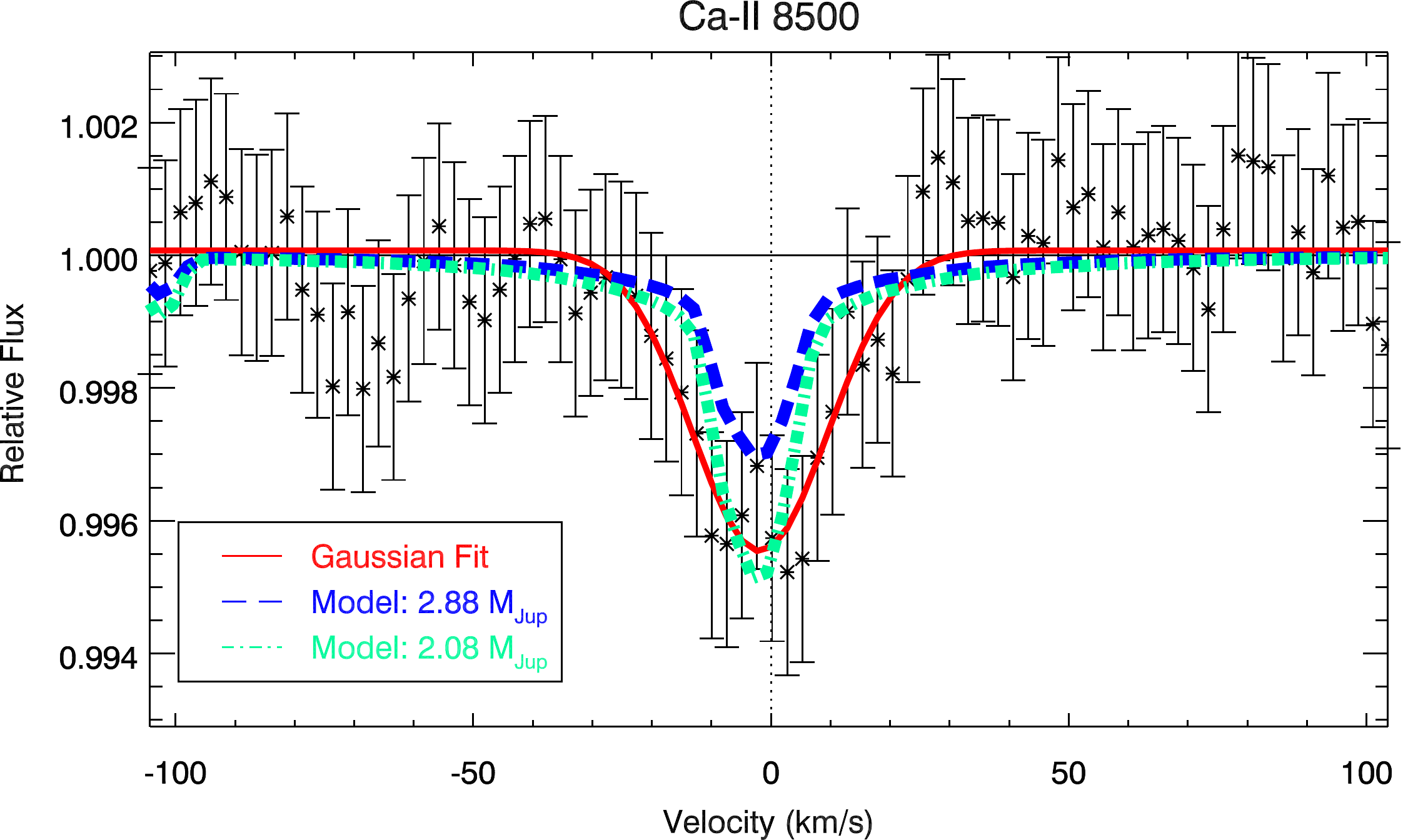} \\
     \includegraphics[width=0.5\textwidth,page=2]{20180616_Model_CaII-8544.pdf} &
     \includegraphics[width=0.45\textwidth,page=1]{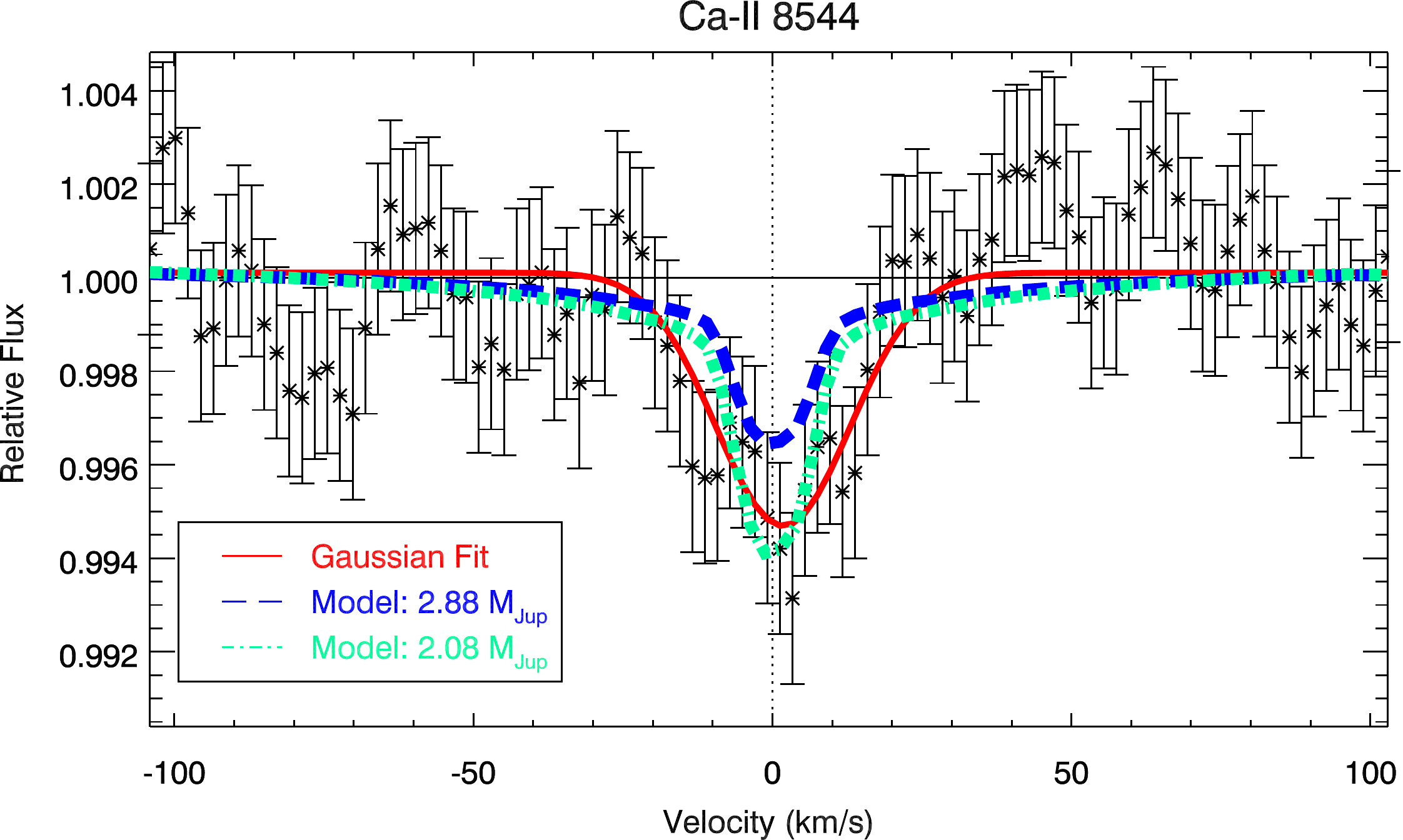} \\
     \includegraphics[width=0.5\textwidth,page=2]{20180616_Model_CaII-8664.pdf} &
     \includegraphics[width=0.45\textwidth,page=1]{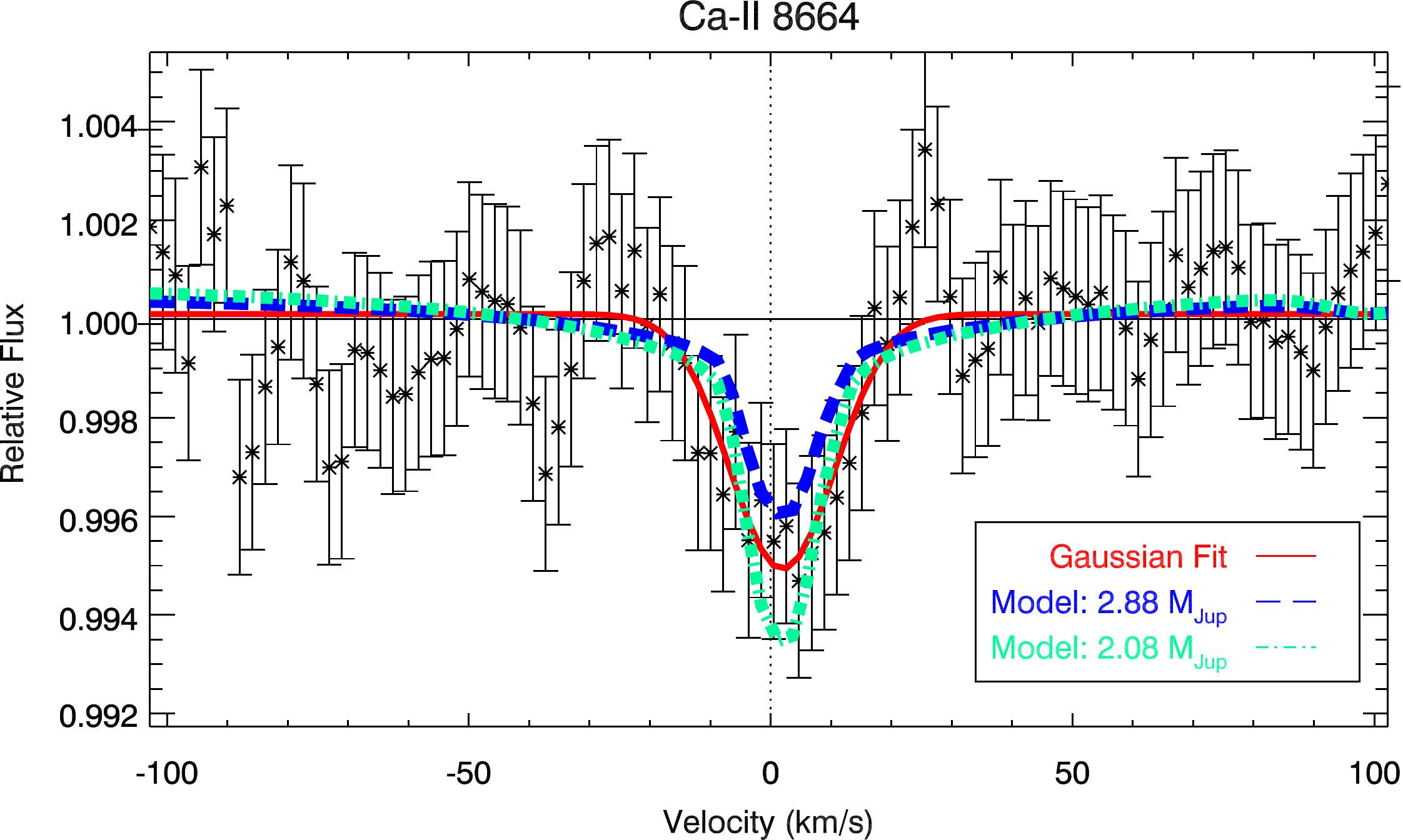} \\
    \end{tabular}
    \caption{Spectra of the ultra-hot Jupiter KELT-9b from the CARMENES spectrograph on the 3.5-m Calar Alto telescope during transit are shown surrounding the H$\alpha$ and ionized calcium (Ca{\sc ii}) infrared triplet lines. The \textbf{left column} shows the final spectrum with absorption from the planet's atmosphere, Doppler shifted due to its changing radial velocity during transit (green-dashed line). The ingress and egress of the transit are shown as solid black lines. The \textbf{right column} shows the H$\alpha$ and Ca{\sc ii} composite absorption profiles in the rest frame of KELT-9b. The red dot-dashed line is a Gaussian fit to the data and the blue dashed curve and green dashed curve are a synthetic atmospheric model with a planetary mass of 2.88\,M$_{Jup}$ and 2.08\,M$_{Jup}$, respectively (see Section \ref{sec:KELT9b_atmo} for more details). The depth of the H$\alpha$ line is 1.02$\pm$0.08$\%$ (compared to the optical transit depth of 0.67$\%$) and indicates the existence of an extended hydrogen envelope around the planet. We find absorption depths of 0.40$\pm$0.5, 0.51$\pm$0.5, and 0.48$\pm$0.6 $\%$ for the Ca{\sc ii} 8500.35 \AA,  8544.44 \AA, and 8664.52 \AA\ lines, respectively.
    }
    \label{fig:absorption}
\end{figure*}

Using these spectra, we found absorption due to H$\alpha$ and the ionized calcium (Ca{\sc ii}) infrared triplet Doppler-shifted with the planet's orbital motion (Figure \ref{fig:absorption}; Table \ref{tb:summary}). For Ca{\sc ii}, the planetary absorption is clearly seen as Doppler-shifted absorption in the spectra at 8500.35 \AA\ line and marginally in the 8544.44 \AA\ line. The planetary absorption profiles for the H$\alpha$ and Ca{\sc ii} can be found in Figure \ref{fig:absorption}. The line profiles for the Ca{\sc ii} lines have been rebinned by two data points to enhance the significance of the detection. H$\alpha$ and all three lines of the Ca{\sc ii} infrared triplet are seen clearly in the absorption profiles.

The planetary line transmission spectra (T$_{\nu}$) are modeled as a Gaussian function
\begin{equation}
 T(\nu) = 1 + \delta e^{\frac{\left(\nu - \nu_{cen}\right)^2}{2\sigma^2}},   
\end{equation}
where $\delta$ is the planetary absorption depth (formally the Gaussian height), $\nu$ is the velocity in the planetary rest-frame, $\nu_{cen}$ is the center velocity of the absorption profile, and $\sigma$ is the Gaussian standard deviation (FWHM = 2.355 $\sigma$). The best-fit model parameters ($h$, $\nu_{cent}$, and $\sigma$) and their errors were found by performing a Differential Evolution Markov Chain Monte Carlo (DE-MCMC) analysis. The same DE-MCMC code is used in the transit modeling program \texttt{EXOMOP} (\citealt{Turner2016b}). In all, we ran 20 chains and 10$^{6}$ links and used the Gelman-Rubin statistic to guarantee chain convergence. The results of the modeling can be found in Table \ref{tb:summary}.

Using this modeling, we find a line absorption depth for H$\alpha$  of 1.02$\pm$0.09$\%$, where the optical transit depth of KELT-9b is 0.68$\%$ (\citealt{Gaudi2017}). The hydrogen atmosphere corresponds to a radius of $\sim$ 1.58$R_{p}$, which extends near the Roche Lobe radius of 1.91 $R_{p}$ (\citealt{Yan2018}). Our H$\alpha$  line depth is within 1$\sigma$ with those found by \citet{Yan2018} and \citet{Cauley2019} of 1.15$\pm$0.05$\%$ and 1.103$\pm$0.010$\%$, respectively. The consistency suggests that the hydrogen envelope around KELT-9b is relativity stable over several years.


Next, we will estimate the hydrogen column density that caused the observed transit signal. Using the analytic equations from \citet{Huang2017}, the number density in the Hydrogen $2{\ell}$ state ($n_{2{\ell }}$) can be estimated using the line center optical depth ($\tau_{0}$)
\begin{equation}
  \tau_{0} = 35 \left( \frac{n_{2{\ell }}}{10^{4} cm^{-3}} \right).
\end{equation}
Furthermore, $\tau_{0}$ can be found by measuring the spectral line width ($\Delta \nu$) of H$\alpha$ 
\begin{equation}
    \Delta \nu = 9.1 \ \text{km \ s$^{-1}$} \left(\ln \tau_{0} \right)^{1/2}.  \label{eq:deltanu}
\end{equation}
Therefore, directly measuring $\Delta \nu$ from the data can provide an estimate on the number density. We fit a line-width ($\sigma$ in the MCMC fit) of 16.3$^{+1.7}_{-1.5}$ km $s^{-1}$. The line width is mainly affected by thermal broadening and the optical depth at the line center. A line width larger than the thermal width indicates that the gas is optically thick. The effect of thermal broadening on the line-width ($\Delta \nu_{T}$) can be measured as 
\begin{equation}
    \Delta\nu_{T} = \sqrt{\frac{8k \ln{2}}{mc^{2}}  }\ T^{1/2} \ f_{0},
\end{equation}
where $T$ is the temperature and $f_{0}$ is the oscillator strength. We find a thermal width of 9.1 km $s^{-1}$ for the H$\alpha$ line where $f_{0}$ = 0.64108 and assuming a temperature of 5000 K. \citet{Huang2017} calculated a mean temperature of 5000 K in the atomic layer of their atmospheric model by balancing photoelectric heating and atomic line cooling. An optical depth of $\tau_{0} \approx 25$ and a number of density of $n_{2{\ell }} = 7.1\times10^{3} cm^{-3}$ are found using the measured line width. Therefore, our data confirm that the the H$\alpha$ line is optically thick.  

We find absorption depths of 0.40$\pm$0.05, 0.51$\pm$0.05, and 0.48$\pm$0.06 $\%$ for the Ca{\sc ii} 8500.35 \AA,  8544.44 \AA, and 8664.52 \AA, lines, respectively (Figure \ref{fig:absorption}; Table \ref{tb:summary}). Using equation \eqref{eq:deltanu} and the measured line widths, we find $\tau_{0} \approx 1.3$. Thus the Ca{\sc ii} lines are optically thick. Our observations confirm the predictions by \citet{Turner2016a} that Ca{\sc ii} should be observable in the upper atmospheres of hot Jupiters. Additionally, our Ca{\sc ii} detection was confirmed recently using independent CARMENES and HARPS-N observations by \citet{Yan2019_caii}. The implications of the Ca{\sc ii} detection are discussed further in Sec. \ref{sec:KELT9b_atmo}.  

Helium (He I; 10830 \AA) has been predicted to exist in hot Jupiter atmospheres (\citealt{Turner2016a}; \citealt{Oklopcic2018}) and has been detected subsequently \citep[eg.][]{Spake2018,Nortmann2018,Allart2018,Allart2019}. We searched our data for He I absorption and found none, in agreement with the results of \citet{Nortmann2018}. Theoretical modeling suggests that, for planets orbiting A stars, photoionization can easily deplete the fraction of helium atoms in the triplet state (\citealt{Oklopcic2019}), thus our non-detection may not be surprising. 

\subsection{Model of KELT-9b's atmosphere} \label{sec:KELT9b_atmo}
In addition to the parametric fits, we have modeled the observed line profiles with synthetic NLTE transmission spectra produced using the spectral synthesis code \texttt{Cloudy} \citep{ferland17}. For the computation of the transmission spectra, we assume a spherically symmetric planetary atmosphere which varies only with altitude. The atmospheric abundance ratio profiles of the first 30 elements, including ions and molecules, as well as the temperature and pressure profiles are adopted from a 1-D PHOENIX model atmosphere as presented in \citet{lothringer18} and \citet{fossati2018}. All other planetary and orbital properties are taken from \citet{Gaudi2017}. Specifically, the mass of the planet is found to be 2.88$\pm$0.84\,M$_\mathrm{Jup}$. We compute the radius at each layer by first determining the pressure scale height, 
\begin{equation}
H_i = \frac{R\,T_i}{\mu_i\,g_i}
\end{equation}
where $R$ is the gas constant, $T_i$ is the temperature of the $i$th layer of the atmosphere, $\mu_i$ is the mean molecular weight of the $i$th layer, and $g_i$ is the planetary gravity evaluated at the $i$th layer (assuming constant planetary mass), and inserting it into
\begin{equation}
r_i = 
\begin{cases}
-H_i\,\ln{(P_i/P_{i-1})} + r_{i-1} & P_i > P_0\\
-H_i\,\ln{(P_i/P_{i+1})} + r_{i+1} & P_i < P_0
\end{cases}
\end{equation}
where $H_i$ is the pressure scale height evaluated at the $i$th layer, $P_i$ is the atmospheric pressure of the the $i$th layer, and $P_0$ is the reference pressure at the planetary surface radius, $r_0=R_P$. We take $P_0$ to be $P_0$=0.01\,$bar$, the pressure in the PHOENIX model at which the equilibrium temperature occurs. This formulation neglects to account for Roche potential effects, which may be important for KELT-9b and will need to be taken into account for more detailed calculations extracting physical information on the properties of the planetary atmosphere from the data.

We map the 1-D model properties onto concentric circles, and calculate the lengths through successive layers of atmosphere along line-of-sight transmission chords as 
\begin{equation}
l_i = c_i - c_{i-1} = \sqrt{2r_ih_i-h_i^2} - \sqrt{2r_{i-1}h_{i-1}-h_{i-1}^2}
\end{equation}
where $l_i$ is the length through layer $i$, $c_i$ is half the chord length along line of sight through layer $i$, $r_i$ is the radius of layer $i$, and $h_i$ is the height down from the top of the atmosphere to layer $i$. These lengths, along with the atmospheric properties of their respective layers, are stacked and entered into \texttt{Cloudy} as the line of sight transmission medium. A schematic of this setup is shown in Figure \ref{fig:schematic} in Appendix \ref{app:SchematicCLOUDY}.

We compute separate transmission spectra with \texttt{Cloudy} for each layer, at a spectral resolution of $R\;=\;120\,000$. The total transmission spectrum of the planet is then the sum of these layer spectra, weighted by the relative area of the stellar disc covered by the respective layers along each line of sight. We assume a non-linear limb darkening function in which we employed limb darkening coefficients extracted on the basis of a library of PHOENIX models \citep{husser2013} as in \citet{salz2019}, and average the wavelength dependence over a $10 \, \AA$ window centered on each line of interest. A more detailed description of the computation of the synthetic transmission spectrum is forthcoming in Young et al. (in prep). 


The right column in Figure \ref{fig:absorption} displays the observed data in comparison to our synthetic profiles. For H$\alpha$ and the Ca{\sc ii} triplet lines, we find that the observed line width is well fit by the synthetic spectra accounting for a turbulent velocity of 3\,km/s, but the line depth is significantly underestimated. The mismatch may be due to the temperature in the model being too low at the line formation layers. We roughly emulate the effect of a higher temperature by reducing the planetary mass in our spectral synthesis by $0.8\,M_\mathrm{Jup}$ (2.08\,M$_\mathrm{Jup}$, a mass that is still within 1$\sigma$ of the derived mass of the planet), thus inflating the atmosphere via reducing the gravity in the scale height calculation. This lower-mass transmission spectrum has an increased line depth and fits the data. 

\begin{figure}[!htb]
   \vspace{1em}
  \sidesubfloat[\textbf{(a.)}]{\includegraphics[width=\textwidth,page=2]{profiles.pdf}}\\
    \vspace{1em}
  \sidesubfloat[\textbf{(b.)}]{\includegraphics[width=\textwidth,page=1]{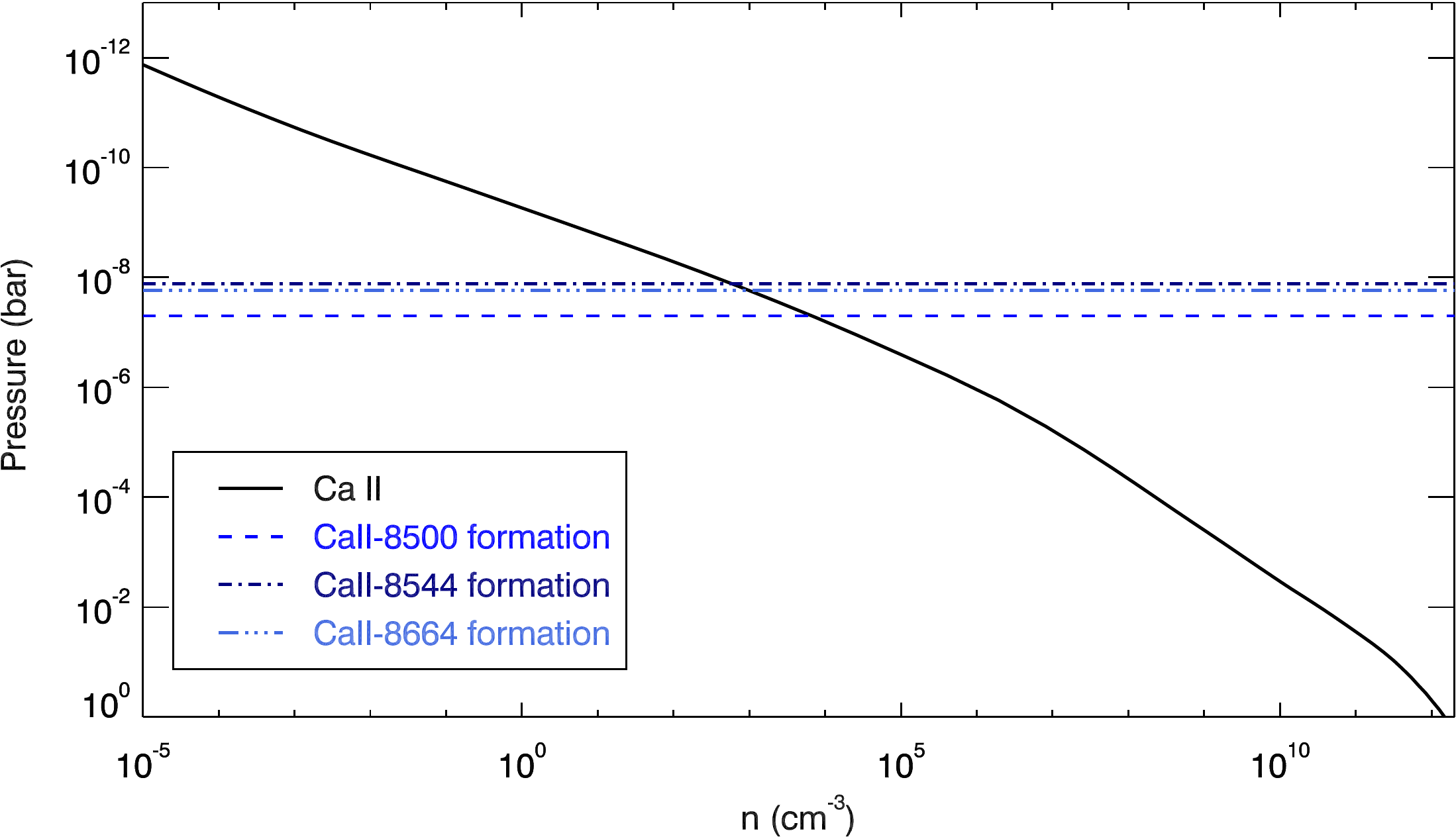}}    
    \caption{Atmospheric profiles for the H$\alpha$ and the Ca{\sc ii} infrared triplet lines derived from the atmospheric model. \textit{Panel a:} Pressure-column density profile of H$\alpha$. H$\alpha$ is formed at a pressure of $\sim$6 nbar with a number density of 28\,cm$^{-3}$ for the 2s state and 86\,cm$^{-3}$ for the 2p state. \textit{Panel b:} Pressure-density profile of three Ca{\sc ii}  infrared lines. The lines are formed between 13 and 50 nbar with number densities between 0.6-- 6.7$\times10^{3}$\,cm$^{-3}$.} 
    \label{fig:profiles}
\end{figure}

Table~\ref{tab:model_params} lists the atmospheric altitude, pressure, and temperature obtained from the PHOENIX model at which the peak of the H$\alpha$ and Ca{\sc ii} triplet lines form considering both models (i.e., planetary mass of 2.08 and 2.88\,$M_\mathrm{Jup}$). The H$\alpha$ line forms at a pressure of about 6\,nbar while the Ca{\sc ii} lines form slightly deeper, at pressures ranging between 13 to 50\,nbar. Figure \ref{fig:profiles} shows the atmospheric density profiles with the pressures at which each line forms.



\begin{table*}[!htb]
    \centering
        \caption{Summary of model atmospheric parameters at line formation altitudes. Column 1: species. Column 2: effective planetary radius at line center. Column 3: atmospheric pressure at effective radius. Column 4: species number density at effective radius. Column 5: species line of sight column density at effective radius. For H$\alpha$ densities, values listed are for n=2 excitation stages.}
    \begin{tabular}{ccccc}
    \hline
    Species                     & Effective Radius & Pressure   & $n$         & $\sigma$ \\
                                & $(R_p)$          & $(nbar)$         & $(cm^{-3})$ & $(cm^{-2})$ \\
    \hline
    H$\alpha$ & 1.28 & 7.9 & 3.6$\times10^{1}$(2s) & 1.8$\times10^{12}$(2s)\\
     &  &  & 1.1$\times10^{2}$(2p) & 5.2$\times10^{12}$(2p) \\
    H$\alpha$ (low mass) & 1.44 & 6.3 &	2.8$\times10^{1}$(2s) &	2.3$\times10^{12}$(2s) \\
     &  &  & 8.6$\times10^{1}$(2p) & 6.9$\times10^{12}$(2p) \\
    Ca{\sc ii} 8500 & 1.21 & 61 & 9.8$\times10^{3}$ & 4.5$\times10^{13}$ \\
    Ca{\sc ii} 8500 (low mass) & 1.32 &	50 & 6.7$\times10^{3}$ &	4.2$\times10^{13}$ \\
    Ca{\sc ii} 8544 & 1.26 & 14 & 6.5$\times10^{2}$ & 3.3$\times10^{12}$ \\
    Ca{\sc ii} 8544 (low mass) & 1.40 &	13 & 6.2$\times10^{2}$ &	4.1$\times10^{12}$ \\
    Ca{\sc ii} 8664 & 1.24 & 20 & 1.4$\times10^{3}$ & 6.9$\times10^{12}$ \\
   Ca{\sc ii} 8664 (low mass) &	1.38 & 17 &	1.0$\times10^{3}$ &	6.7$\times10^{12}$ \\
    \hline
    \end{tabular}
    \label{tab:model_params}
\end{table*}

The 30\% decrease in planetary mass corresponds to a $\sim$1900\,K increase in atmospheric temperature, implying that the temperature between the 1 and 100\,nbar level should be about 8000\,K. Therefore, the observations indicate that the atmosphere should be hotter than predicted in the PHOENIX model. However this does not include the effect that a hotter atmosphere would have on the level populations involved in the line formation, which would probably lead to a further increase in the line strength, hence requiring a slight decrease in the atmospheric temperature. We can therefore conclude that the atmospheric temperature between 10 and 100\,nbar should lie between 6100\,K and 8000\,K. Our finding is in agreement with \citet{Garcia2019}, who also found the need for a lower planetary mass, hence higher atmospheric temperature, to fit the observed H$\alpha$ line profile.

The peak of the H$\alpha$ line, the one forming at the lowest pressure (hence higher altitude) among the four, forms at an altitude of about 1.44 planetary radii (see Table~\ref{tab:model_params}), which is about 0.5 planetary radii below the planetary Roche lobe ($\sim$1.91 R$_{p}$) at the terminator. This shows that the atmosphere at the levels probed by the H$\alpha$ line is not in a hydrodynamic regime and that on KELT-9b this line does not directly probe the escaping upper atmosphere. Instead, the H$\alpha$ line, as well as the other observed Balmer and metal lines, act as atmospheric thermometers enabling one to probe the temperature profile, thus energy budget, which is a key element in controlling the escape. In fact, it is still believed that KELT-9b hosts an escaping upper atmosphere \citep{fossati2018,Garcia2019}, which therefore will require ultraviolet observations to be directly probed (e.g. \citealt{Fossati2010b}; \citealt{Sing2019}).

\section{Conclusions}
In this study, we present high-resolution (R$\sim$94,600) transit observations of KELT-9b using the CARMENES spectrograph on the 3.5-m Calar Alto telescope. We unambiguously detect atmospheric absorption due to H$\alpha$ and the Ca{\sc ii} infrared triplet Doppler-shifted with the planet's orbital motion (Figure \ref{fig:absorption}; Table \ref{tb:summary}). These detections are very robust because different methods used to remove telluric and stellar lines from the data obtained the same final results (Section \ref{sec:sysrem}; Appendix \ref{app:SYSREM}). This detection of Ca{\sc ii} is the first time this species has been observed in KELT-9b's atmosphere and adds to the suite of the 14 already known atmospheric constituents (\citealt{Yan2018}; \citealt{Cauley2019};  \citealt{Hoeijmakers2018,Hoeijmakers2019}). All four absorption lines are found to be optically thick. We also fit the observed absorption line profiles with a synthetic NLTE transmission spectra produced using the spectral synthesis code \texttt{Cloudy} (Section \ref{sec:KELT9b_atmo}; Appendix \ref{app:SchematicCLOUDY}). From this model, we derived the atmospheric altitude, pressure, temperature, and number density where the H$\alpha$ and Ca{\sc ii} triplet lines are formed (Table \ref{tab:model_params}; Figure \ref{fig:profiles}). The Ca{\sc ii} lines are formed at pressures between 13 and 50\,nbar corresponding to an effective radius between 1.32--1.40 R$_{p}$. The H$\alpha$ line forms at an effective radius of 1.44 R$_{p}$ corresponding to a pressure of $\sim$6\,nbar. We also find that the atmospheric temperature between 10 and 100\,nbar should lie between 6100\,K and 8000\,K. Since the H$\alpha$ line is formed well within the planetary Roche lobe (1.91 R$_{p}$; \citealt{Yan2017}), our modeling demonstrates for the first time that on KELT-9b the H$\alpha$ line does not probe the escaping upper atmosphere, though it constraints the atmospheric energy budget which is an important element controlling the escape. In this paper, we show the promise of using individual absorption lines to study the structure of exoplanetary atmospheres.

\acknowledgments
\section*{Acknowledgments}
 We thank Dr. Andrew Ridden-Harper for useful discussions, and the anonymous referee for a thoughtful review.
 
CARMENES was funded by the German Max-Planck-Gesellschaft (MPG), the Spanish Consejo Superior de Investigaciones Científicas (CSIC), the European Union through FEDER/ERF FICTS-2011-02 funds, and the members of the CARMENES Consortium, with additional contributions by the Spanish Ministry of Science, the German Science Foundation (DFG), the Klaus Tschira Stiftung, the states of Baden-W{\"u}rttemberg and Niedersachsen, and by the Junta de Andalucia. 

This research has made use of the Extrasolar Planet Encyclopaedia, NASA's Astrophysics Data System Bibliographic Services, and the the NASA Exoplanet Archive, which is operated by the California Institute of Technology, under contract with the National Aeronautics and Space Administration under the Exoplanet Exploration Program. This research also made use of the NIST Atomic Spectra Database funded [in part] by NIST's Standard Reference Data Program (SRDP) and by NIST's Systems Integration for Manufacturing Applications (SIMA) Program. 


%

\facilities{\textit{3.5-m Calar Alto telescope/CARMENES}; \textit{Exoplanet Archive}}

\software{CARMENES Reduction and Calibration (\texttt{CARACAL v2.01}) pipeline (\citealt{Caballero2016}); \texttt{IDL Astronomy Users Library}; \texttt{Cloudy} (\citealt{ferland17}); \texttt{Spectroscopy Made Easy (SME)} software package (\citealt{Piskunov2017}); We used the following \texttt{Python} packages: \texttt{Astropy},  \texttt{SciPy},  \texttt{NumPy}
}

\bibliographystyle{aasjournal} 
\bibliography{reference.bib} 



\appendix 
\section{\texttt{SYSREM} Tests} \label{app:SYSREM}

\begin{figure}[!htb]
    \sidesubfloat[\textbf{(a.)}]{\includegraphics[width=0.45\textwidth,page=1]{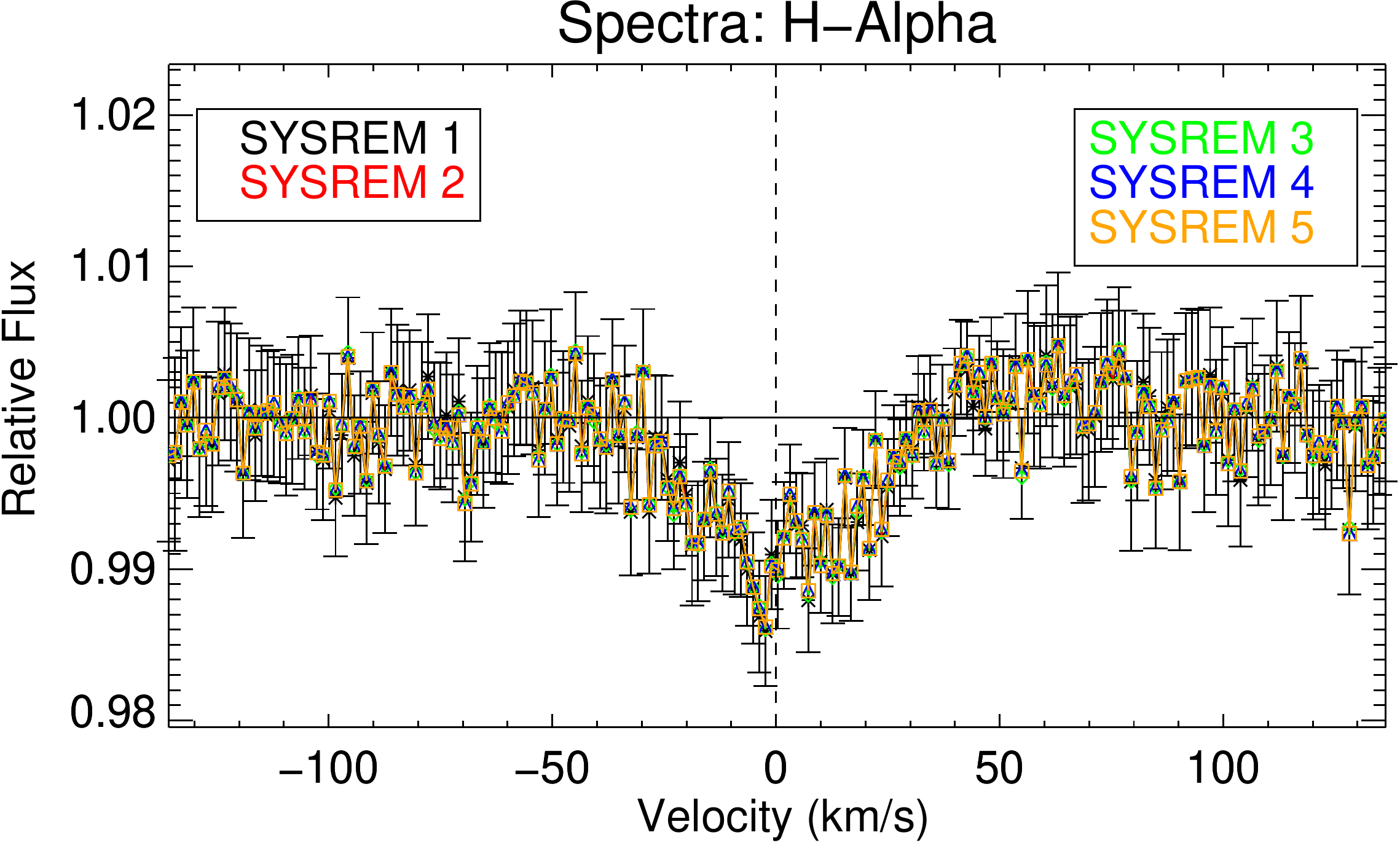}} 
    \sidesubfloat[\textbf{(b.)}]{\includegraphics[width=0.45\textwidth,page=1]{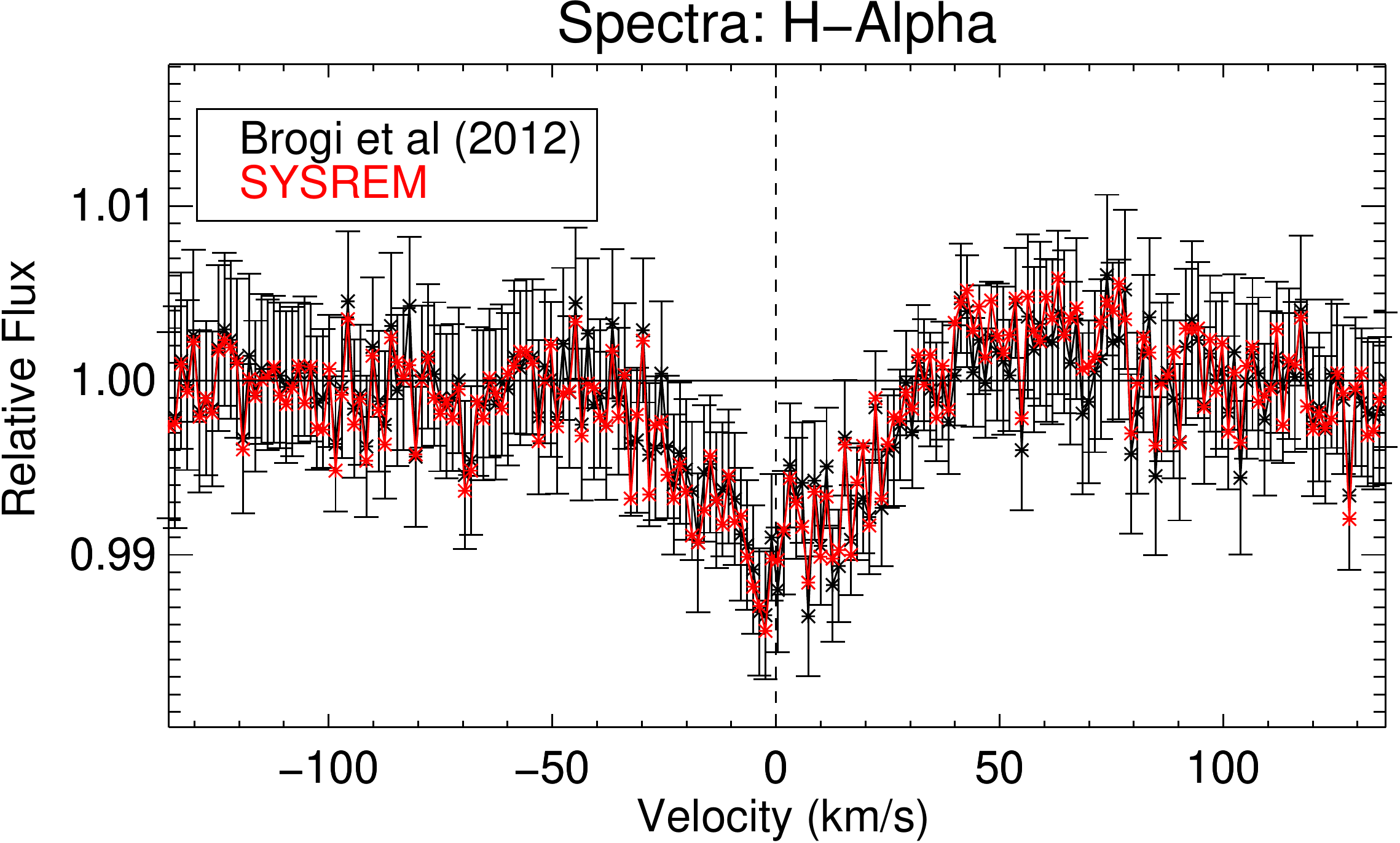}} 
      \vspace{1em}
     \centering \sidesubfloat[\textbf{(c.)}]{ \includegraphics[width=0.45\textwidth,page=2]{SYSREM_Compare_Methods.pdf}}\\
    \caption{\textit{Panel a:} Here we show the H$\alpha$ absorption profiles in the rest frame of KELT-9b derived using a different number of iterations in \texttt{SYSREM}. The error bars are derived from the observational data and only one set is shown for clarity. All the two profiles are identical within the 1$\sigma$ error bars. 
    \textit{Panel b and c:} Here we show the H$\alpha$ and Ca{\sc ii} absorption profiles in the rest frame of KELT-9b derived using \texttt{SYSREM} and the method described in \citet{Brogi2012}. The error bars are derived from the observational data cube and we only show one set of error bars for clarity. The two profiles are nearly identical within the 1$\sigma$ error bars. 
    }
    \label{fig:sysrem}
\end{figure}

In order to efficiently run \texttt{SYSREM}, the procedure has to be performed multiple times. Each iteration removes a different order of systemics due to the telluric and stellar lines. To confirm that running multiple iterations of \texttt{SYSREM} is not removing part of the exoplanet signal we overplot the absorption profiles for H$\alpha$ for different \texttt{SYSREM} iterations in Figure \ref{fig:sysrem}a. All the profiles are identical within the 1$\sigma$ error bars.   

To check the reliability our data reduction with \texttt{SYSREM}, we compared our results to those obtained by using the telluric reduction method described in \citet{Brogi2012} and briefly below. First, the flux of each wavelength is fit as a linear function of the geometric airmass at the time of each observation. Next, the flux of a few of deepest telluric lines were measured over time and the rest of the wavelengths were corrected using a linear regression. Finally, a high-pass filter is applied to each wavelength to remove any remaining low-order structure. Figure \ref{fig:sysrem} shows the comparison of the H$\alpha$ (panel b) and Ca{\sc ii} (panel c) absorption profiles between the two methods. We find consistent results within 1$\sigma$ for both methods. Therefore, if \texttt{SYSREM} is removing part of the planet signal in our data the amount removed is inconsequential on our observed profiles and the resulting physical information of the planet.

\section{Schematic of \texttt{Cloudy} Model} \label{app:SchematicCLOUDY}

\begin{figure}{h}
    \centering
    \includegraphics[width=0.5\textwidth]{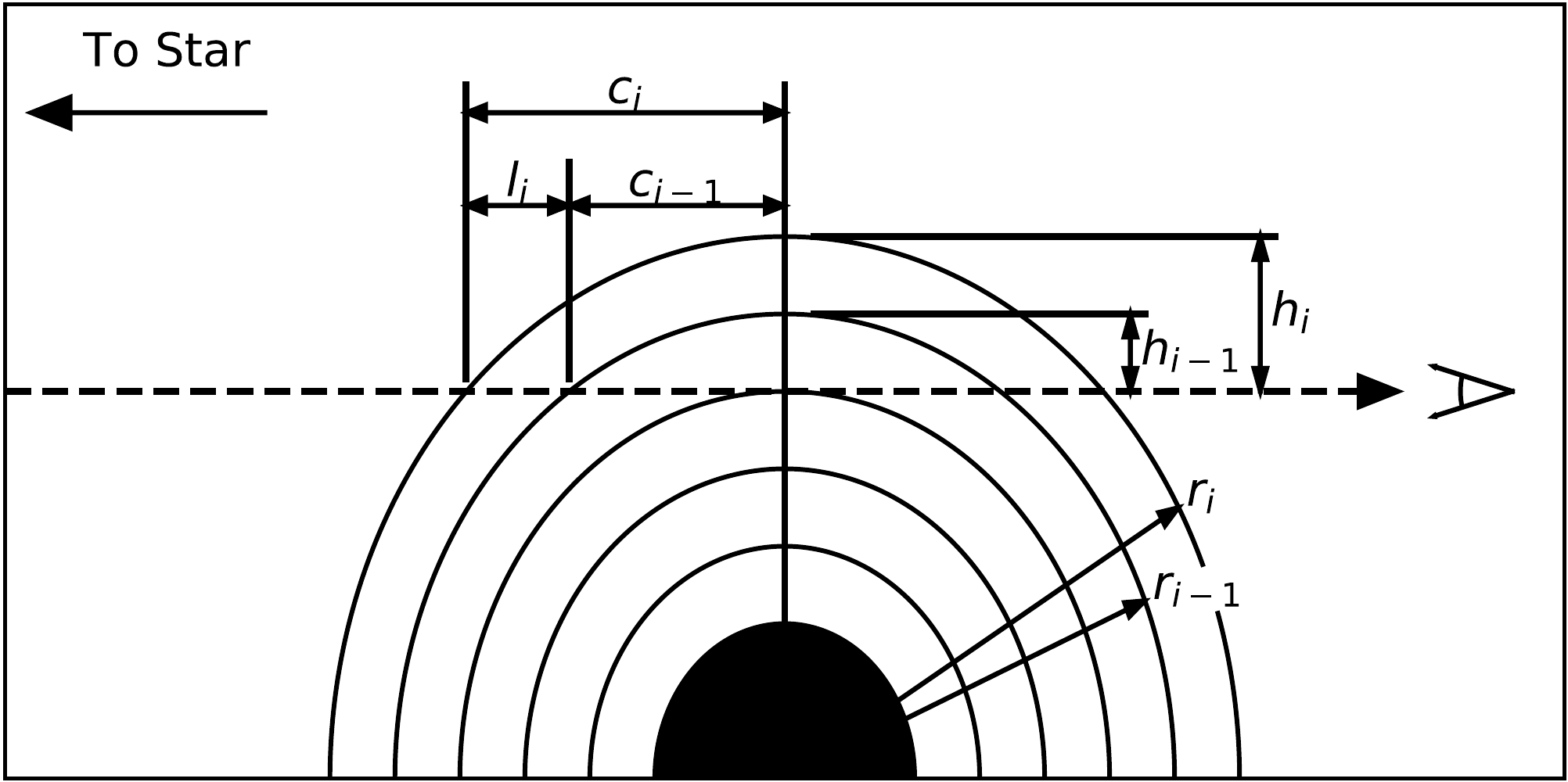}
    \caption{Schematic diagram of transmission chords along the line of sight for transmission spectrum calculation. $c$ is half the chord length along line of sight, $l$ is the path length through a given layer of atmosphere, $h$ is the height down from the top of the atmosphere to the layer, $r$ is the radius of the layer, and the subscripts denote which layer.}
     \label{fig:schematic}
\end{figure}

\end{document}